\documentclass[twocolumn,showpacs,preprintnumbers,amsmath,amssymb]{revtex4}

\usepackage{graphicx}
\usepackage{bm}
\usepackage{dcolumn}

\begin{document}

\title{On a weak solution of Einstein equations for expanding dust}

\author{Sergey~Ph.~Tegai}
\email{ tegai\_s\_f@inbox.ru}
\affiliation{ Institute of Engineer Physics and Radioelectronics,\\
Siberian Federal University,\\
Svobodny 79, Krasnoyarsk, 660041\\
Russia} 

\begin{abstract}{An expanding spherically symmetric dust cloud is considered in a framework of general relativity. Initial conditions leading to a shell-crossing singularity are chosen. The way to construct a weak solution for such a case is proposed. Suggested method consists in cutting off the region containing the shell-crossing and matching the remaining parts of space-time at a thin shell. Junction conditions determine the motion of that thin shell. The singular part of dust stress-energy tensor is nontrivial only after the shell-crossing occurs. Before that the solution coincides with Lemaitre –- Tolman -- Bondi one. A toy model representing an underdensed region in Universe is discussed.}
\end{abstract}


\maketitle

%%%%%%%%%%%%%%%%%%%%%%%%%%%%%%%%%%%%%%%%%%%%%%%%%%%%%%%%%%%%%%%%%%%%%%%%%%%%%%%%%%%%%%%%%%%%%%%%%%%%%%%%%

The Lemaitre -- Tolman -- Bondi (LTB) solution describing the evolution of a dust cloud has recently become quite popular for studying inhomogenities in the Universe. One of the reasons behind that is the opportunity to explain accelerating expansion of the Universe without introducing dark energy. See \cite{Enqvist} for a comprehensive review and clarification of the idea (before 2008) and \cite{AccWithoutDE} for a more recent progress. Another important feature of the LTB space-time is a possibility of describing voids formation (see \cite{Bolejko} including a historical review of voids discovery). And even more cosmological and theoretical applications of that remarkable solution can be found in \cite{Sussman}.

One of the less studied properties of the LTB solution is the formation of shell-crossing singularities (SCS) for certain initial conditions. The cause for it is the intersection of initially different dust layers resulting in diverging and even negative density. The employing of frameworks other then a co-moving one merely brings the metric tensor to a regular form \cite{RegularMetric} but can't remove the singularity because it is in fact a physical but not a coordinate effect. For that reason the initial conditions leading to a SCS are usually avoided even if it seems unfortunate.

The nature of SCS was investigated by different authors \cite{Papapetrou, Tipler, Clarke1997, Nolan1999}  and the conclusion is that it has a different ("weak" or "inessential") type from a shell-focusing singularity and therefore the solution can be extended beyond the SCS.  The first example of such an extension was provided in \cite{Papapetrou} for a rather special case of space-time. Further works \cite{Clarke1997, Clarke1992} suggest the extension to be a weak solution of Einstein equations. In \cite{Nolan2003} such weak solutions are derived treating SCS as a shock wave and using Rankine -- Hugoniot conditions. Unlike the classical solution the weak one is not unique to the future of the shell-crossing singularity even for well posed initial conditions. There is a weak or extended solution which has singular part in stress-energy tensor and there is still a classical solution which is a special case of a weak solution with only regular distributions involved.

Here we introduce another way to find a weak solution employing Israel -- Darmois -- Lichnerowicz junction formalism \cite{Junction}. The idea is to cut out the unphysical regions with negative density and match the remaining parts of the space-time at a thin shell. Of course models with thin shells are nothing new in cosmology and were first studied in \cite{Tomita}. However the important difference is that the thin shell in present work arises from smooth initial conditions.

Israel -- Darmois -- Lichnerowicz matching procedure can be viewed as a consequence of dealing with field equations in a framework of tensor distributions \cite{MarsVickers}. The same is true for relativistic Rankine -- Hugoniot equations which are identical to O'Brien -- Singe conditions while written in general form of energy and momentum conservation across the junction surface \cite{TaubSmoller}. Thus relying on the matching scheme one can expect to get the same results avoiding complicated issues of applying generalized functions to nonlinear theory.

The units with $G = c = 1$ are used throughout.

\section{Joining two Lemaitre -- Tolman -- Bondi space-times at a thin shell}

\begin{figure}[htb]
\begin{center}
\includegraphics[width = 8.2 cm]{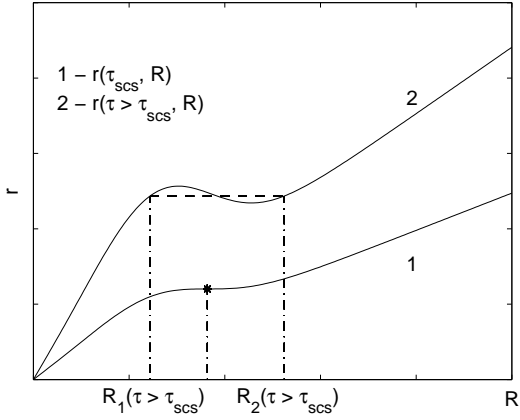}
\end{center}
\caption{\label{Fig1} Radial metric function $r(\tau, R)$ after the formation of SCS (the expanding interval ${[R_1(\tau), R_2(\tau)]}$ should be cut off and replaced with the thin shell), ${\tau_2 > \tau_1}$}
\end{figure}

In comoving coordinates the line element for dust LTB solution is
\begin{equation}
ds^2 = d\tau^2-\frac{{r^\prime}^2(\tau, R)}{1+E(R)}dR^2-r^2(\tau, R)d\Omega^2,
\end{equation}
where ${d\Omega^2 = d\theta^2+\sin^2{\theta}d\phi^2}$. Each layer of the dust is marked with it's own value of radial coordinate $R$. Function ${r(\tau, R)}$ determines the distance to the center of the dust particles with given $R$.  Shell-crossing singularity appears when the layers of dust intersect each other and so the particles with different $R$ start to have the same value of radial function $r(\tau, R)$. This leads to a multi-valued behavior of Misner -- Sharp mass ${m(\tau, r) = m(R) = \left(\dot r^2-E\right)r/2}$ as a function of $r$ \cite{Lasky}. To get rid of the ambiguity one can cut out the layers with multi-valued mass from some point ${R = R_1(\tau)}$ till ${R = R_2(\tau)}$ (see Fig.~\ref{Fig1}). Functions $R_1(\tau)$ and $R_2(\tau)$ should then be found by matching the remaining interior and exterior. 

Not a boundary surface but a thin shell should be used while matching because of the gap between masses $m(R_1)$ and $m(R_2)$. The equation of this shell is specified by ${R = R_1(\tau)}$ for the inner part of dust cloud and by ${R = R_2(\tau)}$ for the outer part. The intrinsic coordinates $\xi^a$ are ${\{\tau, \theta, \phi\}}$ for both. The unit normal has following components
\begin{subequations}
\label{unit_normal}
\begin{equation}
n_\tau = 
-r^\prime_{1, 2}\cdot\left|1+E_{1, 2}-\left(r_{1, 2}^\prime\frac{dR_{1, 2}}{d\tau}\right)^2\right|^{-1/2}
\cdot\frac{dR_{1, 2}}{d\tau},
\end{equation}
\begin{equation}
n_R = 
r^\prime_{1, 2}\cdot\left|1+E_{1, 2}-\left(r_{1, 2}^\prime\frac{dR_{1, 2}}{d\tau}\right)^2\right|^{-1/2},
\end{equation}
\end{subequations}
where index $1$ denotes the values of all functions at ${R = R_1(\tau)}$ while index $2$ does the same for ${R = R_2(\tau)}$.
The only junction condition is continuity on the shell of the first fundamental form with the components
\begin{equation}
g_{ab} \equiv g_{\alpha\beta}e_{(a)}^\alpha e_{(b)}^\beta
\equiv g_{\alpha\beta}\frac{\partial x^\alpha}{\partial\xi^a}\frac{\partial x^\beta}{\partial\xi^b}.
\end{equation}
For interior and exterior parts of the considered dust it is given by
\begin{equation}
d\sigma_{\pm}^2 = \left(1-\frac 1{1+E_{1, 2}}\left(r_{1, 2}^\prime\frac{dR_{1, 2}}{d\tau}\right)^2\right)d\tau^2
-r_{1, 2}^2d\Omega^2.
\label{first_fundamental_form}
\end{equation}
The matching yields
\begin{equation}
\frac{{r^\prime_1}^2}{1+E_1}\left(\frac{dR_1}{d\tau}\right)^2 = 
\frac{{r^\prime_2}^2}{1+E_2}\left(\frac{dR_2}{d\tau}\right)^2;
\end{equation}
\begin{equation}
\label{rr}
r_1 = r_2.
\end{equation}

The first equation has two solutions but only the one with opposite signs provides the required mass discontinuity
\begin{equation}
\label{uu}
\frac{r^\prime_1}{\sqrt{1+E_1}}\cdot\frac{dR_1}{d\tau} = 
-\frac{r^\prime_2}{\sqrt{1+E_2}}\cdot\frac{dR_2}{d\tau}.
\end{equation}
The other solution gives only the classical LTB spacetime.

The full time derivative of (\ref{rr})
\begin{equation}
\dot r_1+r^\prime_1\frac{dR_1}{d\tau} = \dot r_2+r^\prime_2\frac{dR_2}{d\tau}
\end{equation}
together with (\ref{uu}) forms a linear system on $dR_1/d\tau$ and $dR_2/d\tau$. The solution of this system is
\begin{equation}
\label{dR}
\frac{dR_{1, 2}}{d\tau} = \frac{\dot r_{2, 1}-\dot r_{1, 2}}{r^\prime_{1, 2}}\cdot\frac{\sqrt{1+E_{1, 2}}}{\sqrt{1+E_{1, 2}}+\sqrt{1+E_{2, 1}}}.
\end{equation}
The velocity of the thin shell follows immediately from the above equation
\begin{equation}
\label{velocity}
\frac{dr}{d\tau} = \dot r_{1, 2}+r^\prime_{1, 2}\frac{dR_{1, 2}}{d\tau} = 
\frac{\dot r_1\sqrt{1+E_1}+\dot r_2\sqrt{1+E_2}}{\sqrt{1+E_1}+\sqrt{1+E_2}}.
\end{equation}
For marginally bound case this coincides exactly with the shock velocity derived in \cite{Nolan2003} from Rankine -- Hugoniot equations.

\section{A toy model}

Let's consider an LTB space-time with a negative curvature as an example. The solution has the parametric form
\begin{eqnarray} 
r & = & \frac{m(R)}{E(R)}\left(\cosh\eta-1\right),\nonumber\\
\tau & = & \frac{m(R)}{E(R)^{3/2}}\left(\sinh{\eta}-\eta\right)-\tau_0(R).
\end{eqnarray}
With smooth initial conditions one can always choose $r(\tau = 0, R) = R$. With that the solution depends on only two arbitrary functions, say the initial energy density and the initial parameter $\eta$ profiles $\rho_0(R), \eta_0(R)$. All the other functions can be expressed as follows
\begin{equation}
m(R) = \int\limits_0^R4\pi\rho_0(R)R^2\,dR,
\end{equation}
\begin{equation}
E(R) = \frac{m(R)}{R}\left(\cosh{\eta_0(R)-1}\right),
\end{equation}
\begin{equation}
\tau_0(R) = \frac{m(R)}{E(R)^{3/2}}\left(\sinh{\eta_0(R)-\eta_0(R)}\right).
\end{equation}
The square of initial velocity has the form
\begin{equation}
v_0(R)^2 = E(R)+\frac{2m(R)}R = \frac{m(R)}{R}\left(\cosh{\eta_0(R)+1}\right).
\end{equation}

Following \cite{Bolejko} let's take initial conditions at the time of last scattering. It is shown in \cite{Bolejko} that final state of the evolution is less sensitive to the initial density perturbations then to the initial velocity perturbations. So for the sake of simplicity we use a homogeneous initial density profile $\rho_0 = const$. The shape of $\eta_0(R)$ function is chosen in a way that allows formation of a SCS, namely
\begin{equation}
\eta_0(R) = \eta_\infty+(\eta_c-\eta_\infty)\left(1+a_2R^2/R_0^2\right)e^{-a_1R^2/R_0^2},
\end{equation}
where $\eta_c$ and $\eta_\infty$ determine the Friedmann-like behavior of the curvature near the center and at infinity consequently:
\begin{subequations}
\begin{equation}
E(R\rightarrow 0)\approx\frac{4\pi\rho_0}3\left(\cosh{\eta_c}-1\right)R^2,
\end{equation}
\begin{equation}
E(R\rightarrow \infty)\approx\frac{4\pi\rho_0}3\left(\cosh{\eta_\infty}-1\right)R^2.
\end{equation}
\end{subequations}
\begin{figure}[htb]
\begin{center}
\includegraphics[width = 8.2 cm]{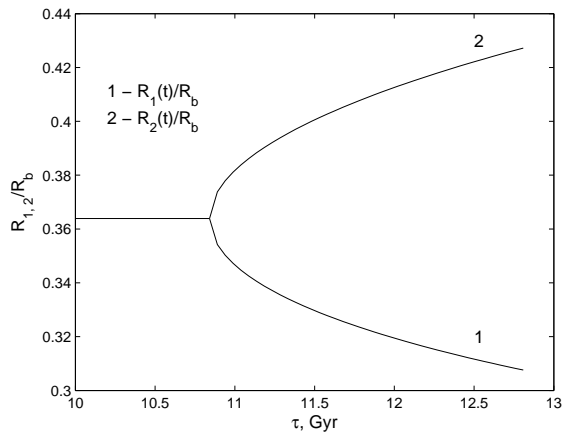}
\end{center}
\caption{\label{Fig2} Numerical solution of junction conditions with $\eta_\infty = 0.01, \eta_c = 0.15, a_1 =  15, a_2 = 20$}
\end{figure}

Now we should solve the system of matching conditions (\ref{rr}) and (\ref{uu}). For numerical calculations it is convenient to work with a new variable ${s = R/R_0}$ The first step is to find a point ${(\tau_{scs}, s_{scs})}$ of the globally earliest occurrence of the shell-crossing singularity. Before the shell-crossing formation the considered matching is trivial and (\ref{rr}) has a unique solution ${s_1 = s_2}$. An arbitrary co-moving surface can be chosen as a junction surface so one can set also ${s_1 = s_2 = s_{scs}}$. But this single root splits into several distinct ones as the shell-crossing occurs. For smooth initial conditions the above splitting first appears when 
\begin{equation}
r^\prime = 0\qquad\mbox{and}\qquad r^{\prime\prime} = 0
\end{equation}
simultaneously. 

Instead of dealing with the DAE system (\ref{rr}) and (\ref{uu}) directly one can use standard methods to solve it's consequence (\ref{velocity}) which is merely an ODE to find the motion of the thin shell. $R_1(\tau)$ and $R_2(\tau)$ are then calculated from ${r(\tau) = r(\tau, R_1(\tau)) = r(\tau, R_2(\tau))}$. The results are displayed at Fig.~\ref{Fig2}.

\begin{figure}[htb]
\begin{center}
\includegraphics[width = 8.1 cm]{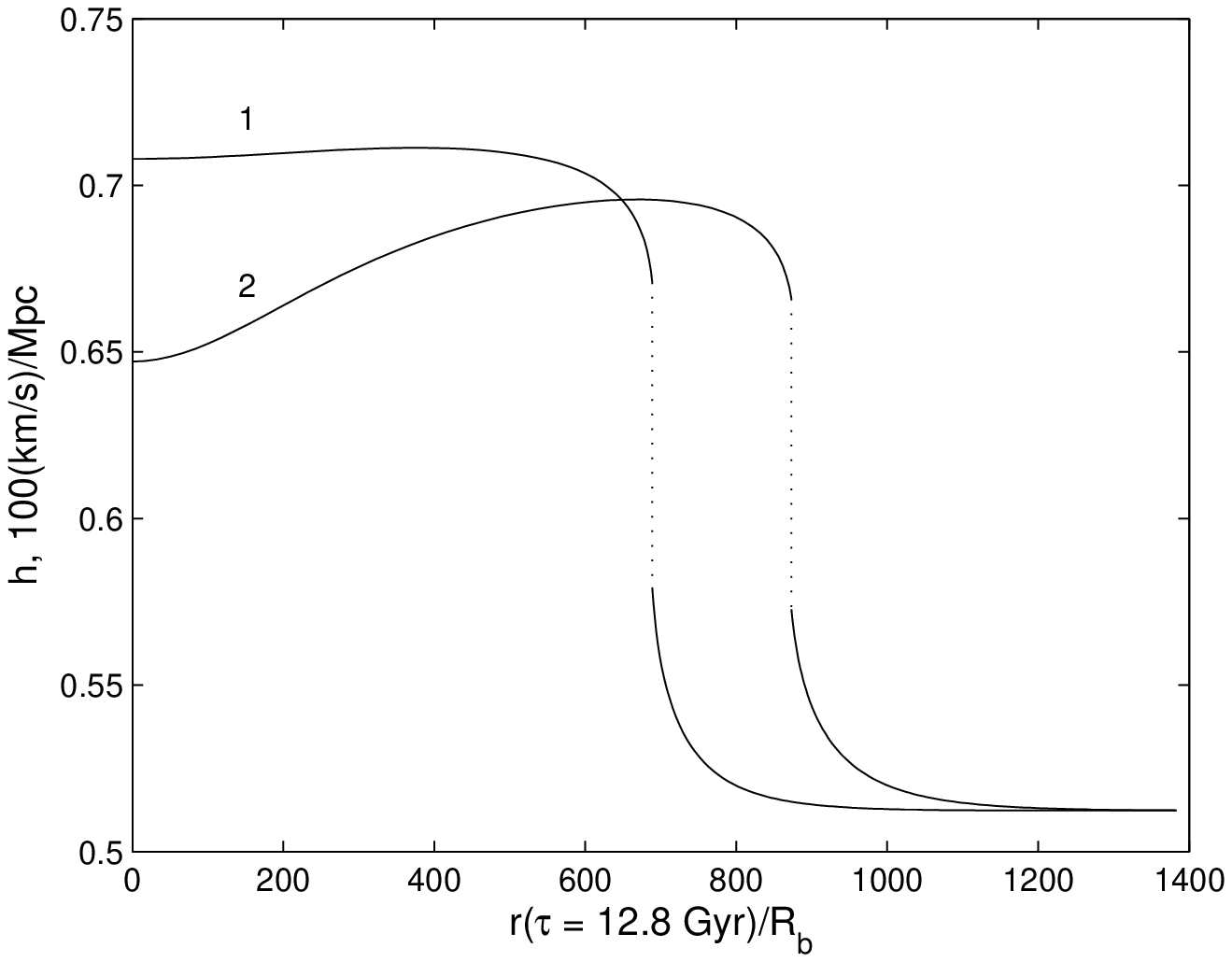}
\includegraphics[width = 8.1 cm]{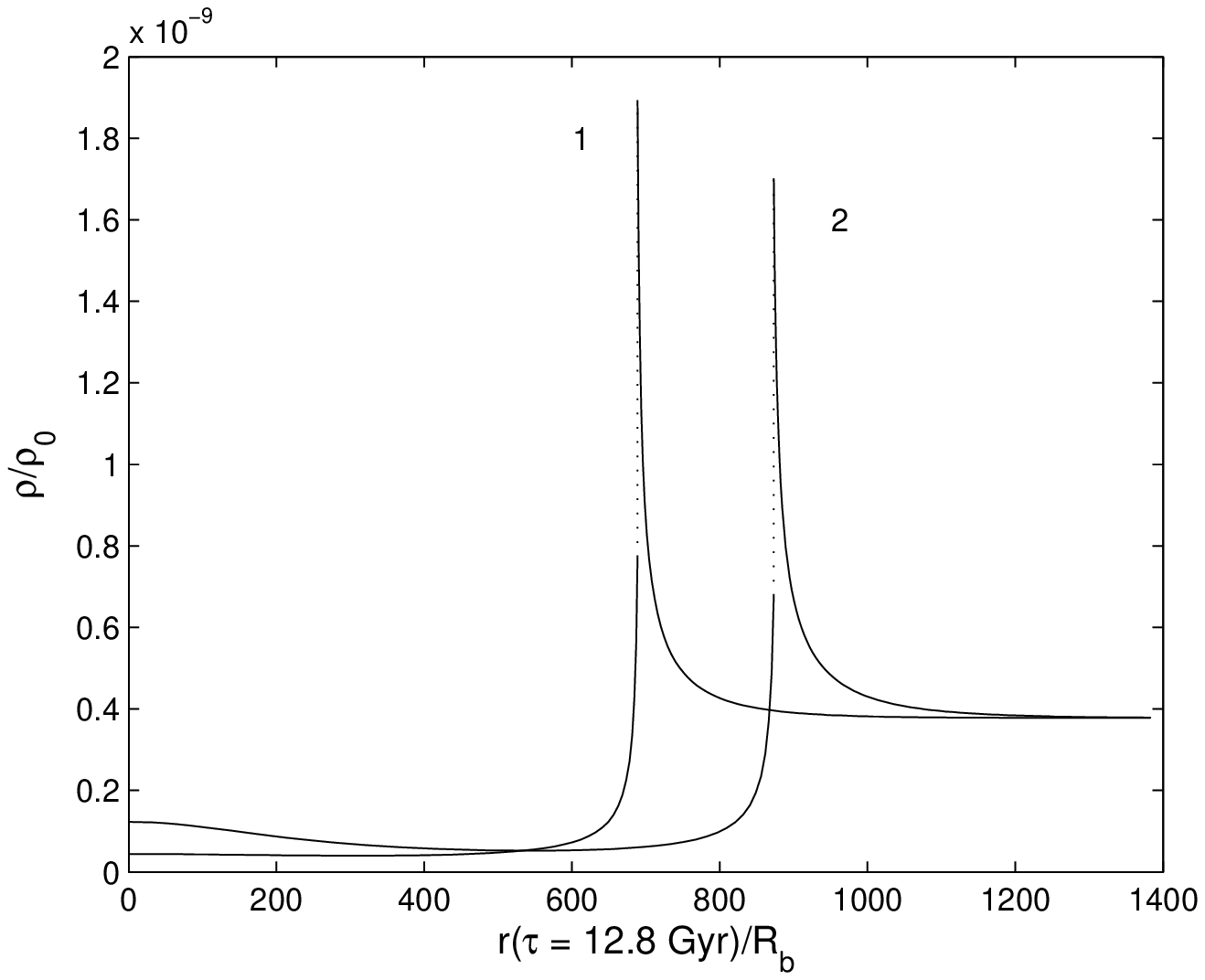}
\end{center}
\caption{\label{Fig3} Hubble constant and density at the present time:\newline
1 --- $\eta_\infty = 0.01, \eta_c = 0.15, a_1 =  15, a_2 = 20$,\newline
2 --- $\eta_\infty = 0.01, \eta_c = 0.09, a_1 =  10, a_2 = 30$
}
\end{figure}

Present time Hubble constant and density of the weak solution are shown at Fig.~\ref{Fig3}. Vertical dotted lines correspond to a positions of the thin shells. Both interior and exterior parts of the weak solution coincide with classical LTB solution but there is a gap between them. 
\section{Discussion}

Stress-energy tensor of the extended solution has the form
\begin{equation}
\label{stress-energy_tensor}
T^{\alpha\beta} = T_{\text{dust}}^{\alpha\beta}+S^{ab}e_{(a)}^\alpha e_{(b)}^\beta\delta_\Sigma({\bf{x}}).
\end{equation}
Here $\delta_\Sigma({\bf{x}})$ is a Dirac's delta function with support on the thin shell. $T_{\text{dust}}^{\alpha\beta}$ is a stress-energy tensor of the dust. $S^{ab}$ is a surface stress-energy tensor related to the extrinsic curvature $K^{ab}$ via the Lanczos equation
\begin{equation}
S_{ab} = -\frac{n^\mu n_\mu}{8\pi}\left([K_{ab}]-[K_a^a]g_{ab}\right).
\end{equation}
Before the SCS occurs the components of $S_{ab}$ are all equal to zero and the space-time is described by Lemaitre -- Tolman -- Bondi solution. After the SCS appears all diagonal components of $S_{ab}$ become nontrivial and can now be expressed in the form of a perfect fluid
\begin{equation}
S_{ab} = (\sigma+p)v_av_b-pg_{ab}
\end{equation}
where $v_a = v_\alpha e_{(a)}^\alpha$ and the 4-velocity of the shell $v^\alpha$ differs from the 4-velocity of the dust. Because of the spherical symmetry the expressions for surface energy density and pressure are simply $\sigma = S^\tau_\tau$ and $p = -S^\theta_\theta = -S^\phi_\phi$. Both of them are positive in our toy model.

Thereby we have constructed a weak solution of Einstein equations with the same initial conditions as classical LTB solution and with positive energy density everywhere including the thin shell. So the main conclusion for that paper is that initial conditions leading to shell crossings should not be forfeited just because of the singularity as it can be replaced with a thin shell. Such initial conditions should be considered equally with any others.

\end{document}